**Idiosyncratic choice bias in decision tasks naturally emerges from neuronal network dynamics**

Lior Lebovich[1]*, Ran Darshan[1,2], Yoni Lavi[3,4], David Hansel[5]† and Yonatan Loewenstein[1,3,6]†

[1]The Edmond and Lily Safra Center for Brain Sciences, The Hebrew University, Jerusalem, Israel

[2]Janelia Farm Research Campus, Ashburn, VA, USA

[3]The Alexander Silberman Institute of Life Sciences, The Hebrew University, Jerusalem, Israel

[4]Code Institute, Dublin, Ireland

[5]Center for Neurophysics, Physiology and Pathology; Cerebral Dynamics, Learning and Memory Lab; CNRS-UMR8119, 45 Rue des Saints Pères, 75270, Paris France

[6]Dept. of Cognitive Sciences and The Federmann Center for the Study of Rationality, The Hebrew University, Jerusalem, Israel

*Correspondence should be addressed to Lior Lebovich (lior.lebovich@mail.huji.ac.il).

†These authors contributed equally: David Hansel and Yonatan Loewenstein.

**Idiosyncratic tendency to choose one alternative over others in the absence of an identified reason, is a common observation in two-alternative forced-choice experiments. It is tempting to account for it as resulting from the (unknown) participant-specific history and thus treat it as a measurement noise. Indeed, idiosyncratic choice biases are typically considered as nuisance. Care is taken to account for them by adding an ad-hoc bias parameter or by counterbalancing the choices to average them out. Here we quantify idiosyncratic choice biases in a perceptual discrimination task and a motor task. We report substantial and significant biases in both cases. Then, we present theoretical evidence that even in idealized experiments, in which the settings are symmetric, idiosyncratic choice bias is expected to emerge from the dynamics of competing neuronal networks. We thus argue that idiosyncratic choice bias reflects the microscopic dynamics of choice and therefore is virtually inevitable in any comparison or decision task.**

*… suppose that there are two equal dates in front of someone gazing longingly at them, unable, however, to take both together. He will inevitably take one of them through an attribute whose function is to render a thing specific, [differentiating it] from its like.*

Abū Ḥāmid Al-Ghazālī (1058 – 1111), The Incoherence of the Philosophers, p. 23[1]

Decision making is the cognitive process of choosing an action among a set of alternatives. Decision making is often studied in experiments, composed of trials, each associated with a single decision. While a decision in a trial is primarily determined by the relevant features of the alternatives, biases are commonly observed[2]. Of specific relevance to this work are *participant-specific* tendencies to prefer one alternative over the others, which we term idiosyncratic choice bias (ICB)[2]. Such biases have been described as early as half a century ago in perceptual discrimination[3–5] and operant learning tasks[6–8].

In discrimination tasks, ICBs interfere with the estimate of perceptual noise. In operant learning experiments ICBs mask the learning behavior. That is why ICBs are typically considered as nuisance. When analyzing choice behavior ICBs are often accounted for by adding an ad-hoc participant-specific bias parameter[4].

Pathological asymmetries can sometimes cause ICBs. For example, patients with visuospatial (or hemispatial) neglect are commonly biased in their responses towards stimuli located in the ipsilesional hemispace[9–11]. In some cases, the specific experimental settings can account for the observed ICBs. For example, in asymmetric settings, in which one alternative is more often associated than others with the correct answer, participants develop a bias in favor of that alternative[12–15].

Sequential effects are also potential contributors to ICBs. In perceptual tasks, a stimulus in a given trial is often perceived as being more similar to the stimuli presented in previous trials[16–18]. In operant learning tasks, correlations between actions and reinforcers bias participants to choose those actions that were previously rewarded[19,20]. These sequential effects can be modelled using linear non-linear regression, in which the bias is a non-linear function of a linear combination of stimuli, actions and the product of actions and reinforcers[21].

One could also interpret ICBs as resulting from sequential effects occurring not during but *prior* to the experimental session[22]. Participants are not *tabula rasa* when entering an experimental session[23]. They come with their own specific history of stimuli and choices. These participant-specific histories can, in principle, result in idiosyncratic biases via long-term sequential effects.

It is generally believed that the neural basis of decision between alternative responses is a competition for higher activity between populations of neurons, each representing a different response. The external input that each population receives is proportional to the relative evidence in favor of the alternative it represents (for review see [24]). An explanation of ICBs in this framework is that they are the result of idiosyncratic asymmetries in the external inputs to the competing populations in the network.

Here we present an alternative explanation. Combining experimental and theoretical results, we argue that substantial ICBs naturally emerge from the dynamics of competing neuronal networks, even in the absence of any asymmetry in their external inputs. According to this explanation, ICBs are expected even when participants are naïve with respect to the task. In fact, our work suggests that ICBs are inevitable unless they are actively suppressed, e.g. by the reward schedule.

## Results

**Choice bias in a bisection discrimination task**

We quantified ICB in the bisection discrimination task depicted in Fig. 1a. In each trial, a vertical transected line was presented on the screen for 1 sec, and participants were instructed to indicate the offset direction of the transecting line (see Materials and Methods). The left panel of Fig. 1b depicts the probability of an 'Up' response as a function of the offset, for three participants. As expected, the probability of a correct response increased with the magnitude of the offset $\Delta L/L \equiv (L^U - L^D)/(L^U + L^D)$, where $L^U$ and $L^D$ denote the lengths of the Up and Down segments of the vertical line. For small offsets, however, the responses differed between the three participants: the blue psychometric curve is shifted to the right of the black curve, whereas the red is shifted to the left.

We considered the choices of the participants in 20 "impossible" trials (1/6 of the trials), in which the line was transected at its midpoint ($\Delta L = 0$). The participant whose psychometric curve is plotted in black in Fig. 1b (left) did not exhibit any significant choice bias, responding 'Up' in 11/20 impossible trials, (p=0.8238, Binomial test). By contrast, the two other participants (red and blue in Fig. 1b, left) exhibited significant choice bias, responding 'Up' in 18/20 and 1/20 of the trials, respectively (p<0.001, Binomial test). Overall, 48% of the participants (n=100) exhibited a significant choice bias (24% significant 'Up', p < 0.05, Binomial test; 24% significant 'Down', p < 0.05, Binomial test). This bias was present despite the fact that we designed the experimental protocol to reduce biases resulting from sequential effects and operant learning (see Materials and Methods and Discussion). Indeed, at the population level, there was no choice bias (p=0.8381, Wilcoxon signed−rank test): the fraction of 'Up' choices in the impossible trials across all participants was 0.505.

To quantify the heterogeneity of these ICBs across the population, we computed, for each of the participants, the difference between the fraction of 'Up' and 'Down' responses in the impossible trials. The distribution of this quantity across the participants is depicted in Fig. 1b (right). Its variance is significantly larger than expected by chance ($p<10^{-6}$, bootstrap, fair Bernoulli process). These results indicate that despite the fact that at the population level, choices in the vertical bisection task were unbiased, the behavior of the individual participants was biased.

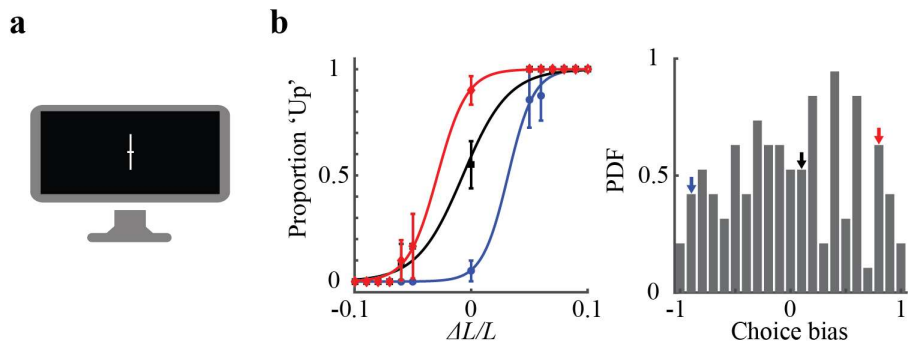

**Figure 1** ICBs in the vertical bisection task. (a) A schematic illustration of the stimulus in a single trial. The participants were instructed to indicate the offset direction of the transecting line. (b) Left: Psychometric curves of three participants. Error bars denote the standard error of the mean (SEM). Curves are best-fit logistic functions. Right: Distribution of the choice bias measure (see Materials and Methods) of all participants ($n = 100$). The biases of the three participants in (b) are denoted in the histogram by arrows of corresponding colors.

**ICB in a motor task**

Next, we constructed a novel motor task, in which ICBs are unlikely to emerge from idiosyncratic sensory asymmetries. In each trial, two adjacent colored dots on a white circular background were displayed on a computer screen (Fig. 2a, see also Fig. S1a). Participants were instructed to drag, as fast as possible, these dots into a central region indicated by a larger black disk. To ensure that the participants would make two temporally-separated reaching movements, we introduced a 1.1 sec delay after the completion of the dragging of the first colored dot (Materials and Methods). The task was presented to the participants as a motor-speed task, in which faster movements are more rewarded (see Materials and Methods). However, the behavioral parameter that we were interested in was the order in which participants chose to execute the two dragging movements. In that sense, this paradigm is a binary, implicit, decision-making task in a setting which is symmetric with respect to the task. In this task, choice bias manifests as a tendency to choose to drag one of the two dots first more often than expected by chance. Each participant was presented with 10 different pairs of dots of different colors and locations. Each of these pairs was presented 20 times in a pseudorandom order. The ICB of a participant for a given pair of dots was defined as the difference between the fraction of trials, in which she chose the clock-wise dot first and the fraction of trials, in which she first chose the counter-clockwise dot. This allowed us to measure 10 different ICBs (one for each pair) for each participant.

Figure 2b depicts the distribution of choice biases across the participants for the pair of dots plotted in Fig. 2a. There was no systematic choice bias at the population level ($p=0.41$, Wilcoxon signed−rank test). Nevertheless, 65% of the participants exhibited significant ICB for this pair (35% significant preference towards choosing the clock-wise dot first and 30% significant preference in favor of choosing the counter-clockwise dot first; $p < 0.05$ Binomial test).

Considering all 10 pairs, for 8/10 of the pairs we found no consistent choice bias across the population (p>0.05, Wilcoxon signed−rank test, not corrected for multiple comparisons). These 8 pairs were considered for further analysis (see Fig. S1b for all pairs). For each of these, the variances of the bias distributions were larger than expected by chance (Fig. 2c; $p>10^{-6}$, bootstrap, fair Bernoulli process), indicating that participants in this task exhibited significant ICBs for each of these pairs.

It should be noted that the distributions of biases in this motor decision task was different from the one we found in the bisection task. Although both were broader than expected by chance, the distribution was narrower in the bisection task than in the motor task (compare Figs. 1b, right, and 2c; standard deviations are 0.55 and 0.68, respectively, $p<10^{-4}$, Shuffling).

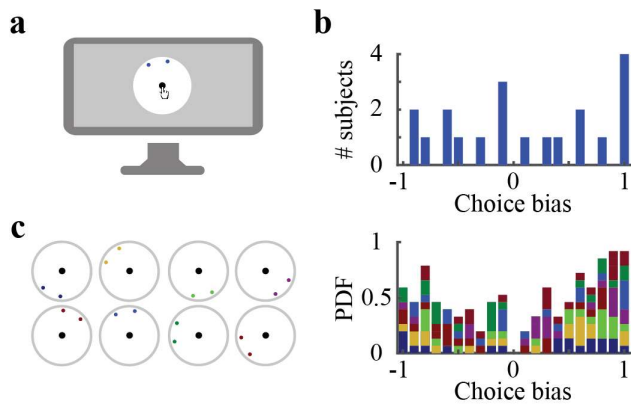

**Figure 2:** Choice bias in the motor task. (a) A schematic illustration of the stimulus in a single trial. The participants were instructed to use the mouse to drag the two dots to a center circle as fast as possible. We used 10 different pairs of dots, each appearing 20 times (see also Fig. S1a). (b) Distribution of ICBs over all participants ($N = 20$) for the pair of dots depicted in (a). Positive (negative) bias corresponds to a preference towards choosing the clockwise (counterclockwise) first. (c) Right, Distribution of choice biases across the population for the 8 pairs of dots (Left, color-coded) devoid of a consistent choice bias (see also Fig. S1b).

**Choice bias and the Drift Diffusion Model**

The two experiments described in Figs. 1 and 2 demonstrate ICBs in two-alternative decision tasks that are symmetric, when judged by the population-average behavior. To understand the neural basis of these ICBs and the determinants of the shape of their distributions, we constructed a simplified neuronal network that models choice behavior in the bisection task. The network consists of two populations of neurons representing 'Up' and 'Down' choices, denoted by '*U*' and '*D*', respectively (Fig. 3a, left). Each population comprises $N/2$ independent Poisson neurons, such that the spike train of each neuron in a trial is an independent homogeneous Poisson process. The firing rate of each neuron is an exponential function of the sum of offset-dependent and offset-independent inputs (Eq. 1 in Materials and Methods). The offset-dependent input is a linear function of the offset of the bisecting line $\Delta L$ (Fig. 3a, bottom; see Materials and methods), such that the firing rates of the *U* neurons increase with $\Delta L$, whereas that of the *D* neurons decrease with $\Delta L$ (Fig. 3a, right). A negative offset ($\Delta L < 0$) decreases the firing rate of the *U* neurons and increases that of the *D* neurons. The offset-independent input models the heterogeneity between the neurons by assuming that each neuron receives an additional input, different for each neuron and drawn from a normal distribution, constant over time and between trials. As a result, the firing rates of the neurons are drawn from log-normal distributions (as is observed in the cortex[25,26]), whose parameters depend on the offset (orange and pink distributions in Fig. 3a). In the absence of an offset ($\Delta L = 0$), the firing rate distributions of the two populations are the same (blue distribution in Fig. 3a, right).

Decision in our model depends on the cumulative number of spikes, $n^U(t)$ and $n^D(t)$, emitted by populations *U* and *D* up to time $t$ in a trial. A decision is made at time $t^*$, at which the absolute value of the difference in the numbers of spikes, $|\Delta n(t^*)| = |n^U(t) - n^D(t)|$, reaches a given threshold, $\theta$, for the first time. The decision is 'Up' if $\Delta n(t^*) = \theta$, whereas it is 'Down' if $\Delta n(t^*) = -\theta$ (Fig. 3b).

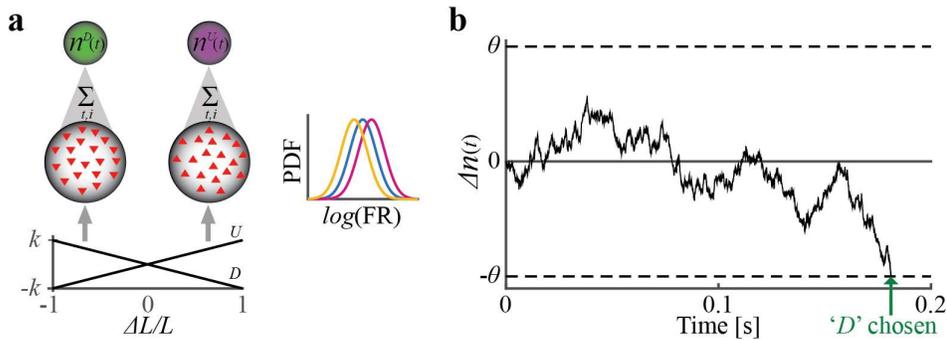

**Figure 3:** The DDM. (a) Schematic illustration of the network. It consists of 2 large populations of independent Poisson neurons (Center), receiving stimulus-selective input (Bottom, direction of triangle denotes selectivity to $\Delta L$) and emitting spikes which are accumulated (Top). Right, the stimulus-dependent distribution of firing rates. In the absence of offset ($\Delta L = 0$), the rate of 'U' and 'D' neurons are drawn from the same distribution (blue curve). When the upper segment of the line is longer ($\Delta L > 0$), neurons in population 'U' increase their firing rates (pink curve), whereas neurons in population 'D' decrease their firing rates (orange curve). Note that the lognormal distribution of rates is equivalent to normal distribution of log-rates. (b) Example trial. The absolute value of the difference in spikes counts is accumulated over time, until the threshold is reached. The decision corresponds to the 'winning' population, here 'D'.

Our model is a microscopic version of the pure Drift Diffusion Model of decision making (DDM) in the free-response paradigm[24,27–34]. In the DDM, noisy evidence in favor of each of the alternatives is accumulated throughout the trial and its difference is represented by a dynamic variable. A decision is made when the absolute value of this variable reaches a predefined threshold and the identity of the decision is determined by the sign of this variable at the time of the decision. In our model, the accumulated evidence in favor of an alternative is the number of spikes emitted by the corresponding population of neurons.

The psychometric curve of an example network is depicted in Fig. 4a (left; black). Because of the dependence of the firing rate distributions on $\Delta L$, the larger $\Delta L$, the more likely it is that the network would choose 'Up'. However, the outcome of this decision process is not deterministic. Because of the Poisson nature of the spiking, $\Delta n(t)$ occasionally reaches the threshold that is incongruent with the stimulus, resulting in an error. More generally, because the firing of the neurons is stochastic, the psychometric curve is a smooth sigmoidal function of $\Delta L$ rather than a step function. Note that in the black psychometric curve of Fig. 4a, the network's perceptual decision in the "impossible trials" ($\Delta L = 0$) is approximately at chance level. Thus, this particular network does not exhibit a substantial choice bias.

The black psychometric curve in Fig. 4a (left) was obtained for a particular realization of the network. The red and blue lines in Fig. 4a (left) depict the psychometric curves of two other realizations. Despite the fact that the three networks were constructed in the same way, i.e., by randomly drawing the firing rates of the neurons from the same distributions, the red psychometric curve is shifted to the left whereas the blue is shifted to the right, relative to the black one. Thus, in contrast to the "black" network, the "red" and "blue" networks exhibit ICBs in favor and against responding 'Up'. We mathematically derived the distribution of ICBs in the DDM and its dependence on the model parameters (see Materials and Methods). This is depicted in Fig. 4a, center. These results demonstrate that a wide distribution of ICBs naturally emerges in a symmetric setting in this standard decision model.

Observing a wide distribution of ICBs in networks comprising of a small number of neurons is easy to understand. Let us consider a model composed of only two Poisson neurons (Eq. 1 in Materials and Methods), one representing choice 'Up' and the other representing choice 'Down'. Even in the impossible trials in which the firing rates of the two neurons are independently drawn from the same lognormal distribution, the actual firing rates of the two neurons will, in general, differ. In some realizations the firing rate of the $U$ neuron will be higher than that of the $D$ neuron, whereas in others, it will be lower. Choice, determined by the first threshold reaching of the accumulated difference in the number of spikes fired by the neurons, will more often be congruent with the neuron whose firing rate is higher. However, because the firing of spikes in the model is a Poisson random process, in some trials decision will be incongruent with the difference in the firing rates. In summary, in this two-neuron network, ICB is the result of the interplay between the Poisson noise and the heterogeneity in the (two) firing rates. The former decreases the bias, whereas the latter increases it.

Emergence of ICB is thus expected in small decision-making networks. However, it is not immediately clear why biases are also observed in our DDM model, which involves a large number of neurons. In our simulations, the difference between the *population averaged* firing rates of the $U$ and $D$ neurons is vanishingly small. This is because in large networks, this difference is of the order of $1/\sqrt{N}$, where $N$ is the number of neurons, and thus goes to zero when $N$ is large (as in Fig. 3, $N = 200{,}000$). Thus, one may expect that in our DDM model, this heterogeneity in firing rates should not play a significant role in the decision process. One should note, however, that the sensitivity of the decision-making network to the difference in the average firing rates increases in proportion to $\sqrt{N}$. This is because the trial-to-trial Poisson-driven fluctuations decrease with the number of the neurons $N$. This increases the sensitivity of the choice process to small heterogeneities in the firing rates and by that, results in substantial ICBs even in large networks. As a matter of fact in the limit of infinite $N$, the distribution of ICBs remains broad and becomes independent of $N$ (Materials and Methods).

Unlike network size, the decision threshold has a large effect on the magnitude of choice bias. This is depicted in Figs. 4a-c, where the psychometric curves of three networks, only differing in the value of the decision threshold, are plotted (left). The larger the threshold, the steeper is the psychometric curve. This is because the time it takes the network to reach a decision increases with the threshold (Compare right panels in Figs. 4a-c). Thus, a larger threshold results in the integration of spikes over longer durations before a decision is made. Therefore, decision outcomes are less sensitive to the Poisson noise. On the other hand, network heterogeneity is independent of decision time. Because the magnitude of the choice bias is determined by the interplay of the Poisson noise and networks heterogeneity, the larger the threshold is, the broader will be the distribution of ICBs (Figs. 4a, 4b and 4c, Center).

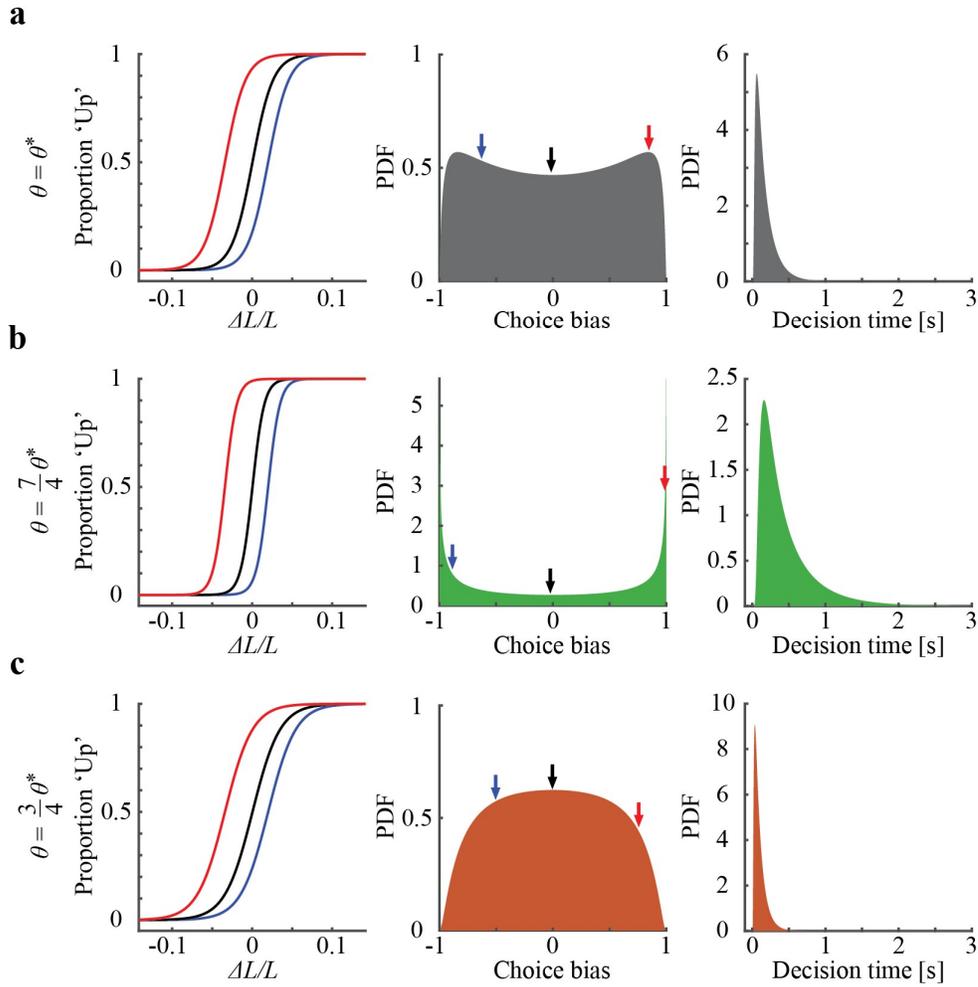

**Figure 4:** ICBs in the DDM. The psychometric curves (Left), ICB distribution (Center) and distribution of decision times (Right) for intermediate (a), high (b) and low (c) thresholds, (See: Materials and Methods), where $\theta^* = 291$. Left: Psychometric curves depicting 3 different networks (Eq. 2 in Materials and Methods). *Center*: ICB distribution (Eq. 3 in Materials and Methods). Right: Distribution of decision times; (Eq. 4 in Materials and Methods).

**Choice bias in a recurrent spiking network**

In our DDM, both the Poisson firing of the neurons and the heterogeneity in their firing rates, the ingredients of our mechanism for the ICB, were introduced *ad hoc*. Therefore, we investigated whether the results derived in the latter framework also hold for more realistic recurrent spiking network models of decision making, in which noise and heterogeneity emerge from the non-linear collective dynamics of the network.

The model we studied comprises of 32,000 excitatory and 8,000 inhibitory Leaky Integrate and Fire (LIF) neurons (Fig. 5a; see Materials and Methods for details). All neurons receive a feedforward input, which is selective to the stimulus. For half of the neurons (U neurons), this input linearly increases with $\Delta L$, whereas for the other half, (D neurons), it is a decreasing function of $\Delta L$. When the two segments are of equal length (impossible trials), the U and D neurons receive the same stimulus-dependent feedforward input. All neurons are recurrently connected in a random and non-specific manner, i.e. independent of the selectivity properties of the pre and postsynaptic neurons. The competition between the U and the D neurons is mediated by an additional set of connections, which are functionally specific, less numerous but stronger than the unspecific ones. We investigated the dynamics of this model by performing numerical simulations (See Materials and Methods).

Because of the strong recurrent connections, the network operates in the balanced regime[35], in which strong inhibition compensates for the strong excitation. The activity of the neurons in the network, in the absence of stimulus-related input (spontaneous activity), exhibits Poisson-like temporal variability of spike timing (Fig. 5b). Firing rates are heterogeneous across neurons and are approximately log-normally distributed[36] (Fig. 5c) as also observed in the cortex[25,26].

Before a stimulus is presented, the population average firing rates of the U and D neurons are similar (Fig. 5d). In response to the stimulus ($t = 0$), the neurons increase their firing rates. Because of the competition induced by the specific inhibitory connectivity between these populations, the time courses of the activities of the *U* and the *D* neurons are different. The decision process in this model is determined by the relative difference in the average firing rates of the excitatory neurons of the *U* and the *D* populations. The decision is made when this difference is larger than a fixed threshold, in congruence with the more active population (Fig. 5d; see also Materials and Methods).

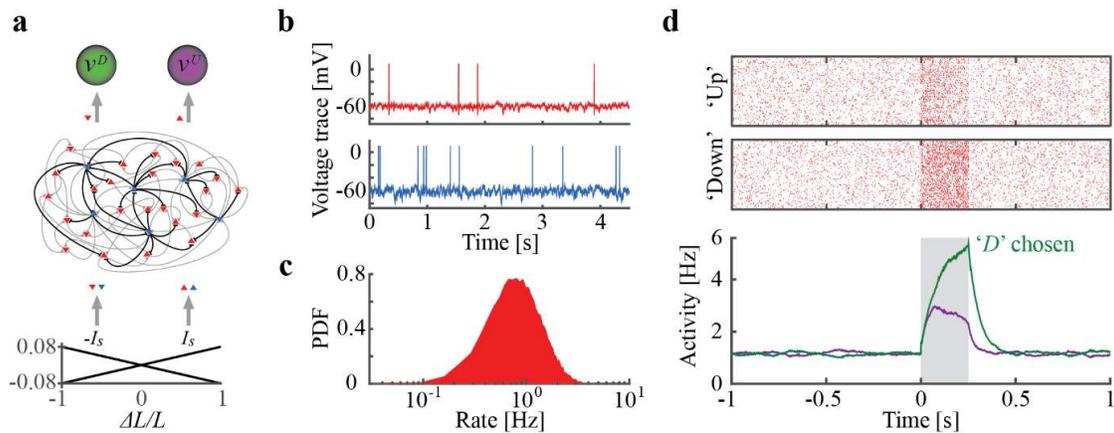

**Figure 5:** Spiking network model. (a) Schematic illustration of the network architecture. The network (top) consists of recurrently connected excitatory (red) and inhibitory (blue) LIF neurons receiving stimulus-selective feed-forward input (bottom, direction of triangle indicates the selectivity). (b) Spontaneous activity of example excitatory (red) and inhibitory (blue) neurons in the network. (c) Distribution of the spontaneous firing rates of the 32,000 excitatory neurons in the network. (d) Raster plot (10% of the neurons, top) and average firing rates (bottom) of the excitatory *U* (purple) and *D* (green) neurons, in response to a stimulus (gray region). The decision is made when the relative difference between the firing rates of the two populations crosses the threshold (after which the feed-forward input ceases).

Fig. 6a (left, black) plots the psychometric curve of the network depicted in Fig. 5. When the magnitude of $\Delta L$ is large, the perceptual decision of the network is almost always correct; as $\Delta L$ decreases, the error rate increases. Considering the "impossible trials" ($\Delta L = 0$), in which the two segments are of equal length, the network's perceptual decision is approximately at chance level. However, a different realization of the connectivity matrix will result in a different psychometric curve. The red and blue curves in Fig. 6a, left, depict the behavior of two additional networks, each corresponding to a different realization of the connectivity matrix. In contrast to the "black" network, the "red" and "blue" networks exhibit substantial choice bias.

To estimate the distribution of ICBs in our recurrent network model, we simulated 200 networks, which only differed in their realizations of the connectivity matrix. We computed the ICB of each network from choice in 500 "impossible" trials. The central panel in Fig. 6a, center, depicts the distribution of these ICBs across the 200 networks. It is significantly wider than expected by chance ($p<10^{-6}$, bootstrap, fair Bernoulli process).

In the DDM, the value of the threshold controls the shape of the distribution of biases because it affects the average decision time. We studied the determinants of the shape of the bias distribution in our recurrent network model. Simulations revealed that this shape depends on the strength of the selective inhibitory synapses $g$ (Materials and Methods section). This is because $g$ strongly modulates average decision time. As depicted in Fig. 6b, decreasing $g$ slows down decision (compare Fig. 6b, right to Fig. 6a, right) and results in steeper psychometric curves (compare Fig. 6b, left and Fig. 6a, left) and a wider, more convex, distribution of ICBs (compare Fig. 6b, center and Fig. 6a, center). Similarly, increasing $g$, results in faster decision, shallower psychometric curves and a more concave distribution of ICBs (Fig. 6c).

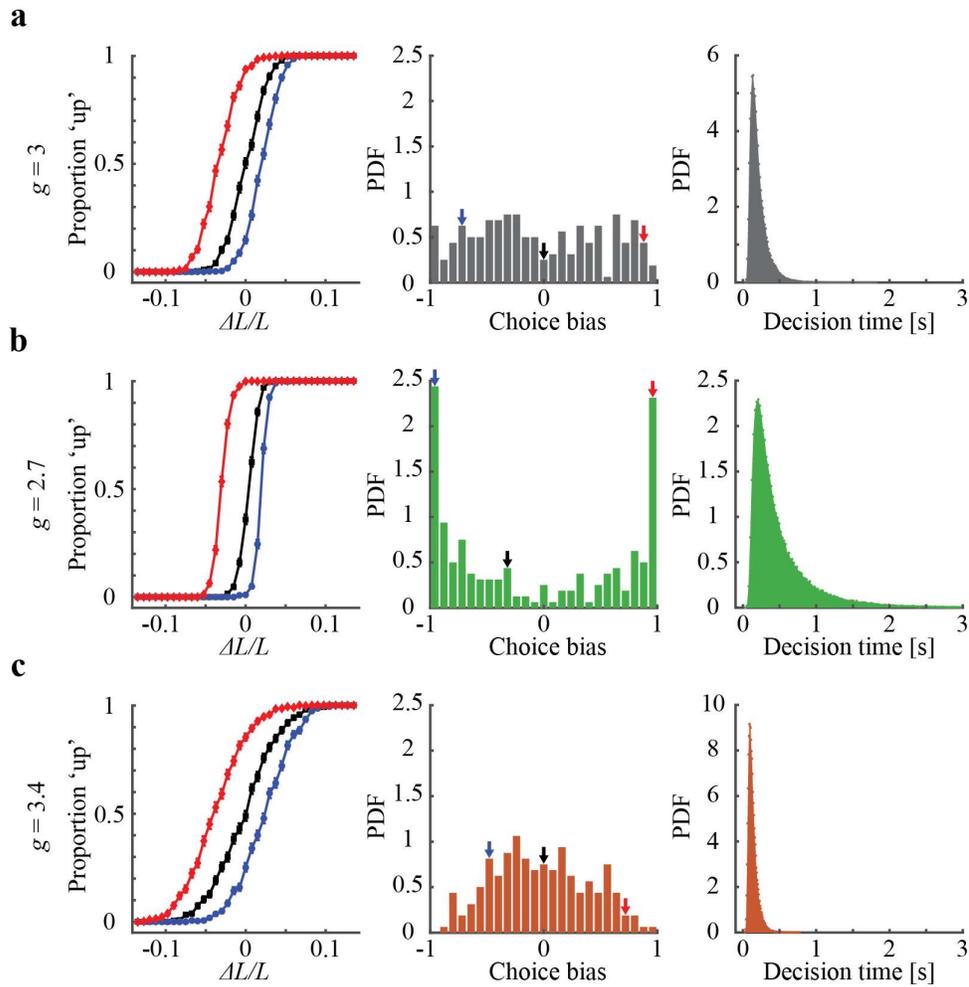

**Figure 6:** Choice bias in the recurrent network model. (a-c). Left: Psychometric curves for 3 different realizations of network connectivity. Each point is an average over 500 trials. Error bars correspond to SEM. Center: Distribution of ICBs for 200 networks. Arrows correspond to specific example networks of the psychometric curves (color coded). Right: Distribution of decision times. The strength of the selective inhibition is (a) $g = 3$; (b) $g = 2.7$; (c) $g = 3.4$.

**Discussion**

We experimentally investigated human choice bias in a discrimination task and a motor task. While in both tasks there was no significant bias in favor of one of the responses at the population level, *individual* participants exhibited a preference towards one of the responses (ICBs). We also investigated two computational models of decision making, one based on the DDM and the other on a recurrent network of LIF spiking neurons. Both models exhibit broad distributions of ICBs across statistically-identical networks. The shape of these distributions reflects the interplay between fast noise and heterogeneities in the fine structure of the network. Our theoretical results show that ICBs can emerge naturally from the dynamics of decision making neural circuits.

**The temporal-scale of stochasticity**

In the two models that we have investigated, ICBs emerge from the interplay of two sources of stochasticity: (1) Heterogeneity in the neuronal firing rates; (2) Stochasticity in the timing of action potentials. The first source of stochasticity is the same in all trials and therefore, we refer to it as *frozen* or *quenched* noise. By contrast, stochasticity in the timing of action potentials differs between trials and therefore we refer to it as *fast* noise. In the DDM, both types of stochasticity are explicitly introduced *ad hoc*. In the recurrent spiking model, they emerge from the randomness of connectivity and the collective dynamics of the network.

Stochastic processes in cortical networks occur at multiple time-scales[37]. Incorporating additional time-scales to the models will not qualitatively affect the results, as long as the contributions of these additional sources of stochasticity are on the order of $1/\sqrt{N}$ (where $N$ is the size of the network). To identify stochasticity at minutes' time-scale, we tested whether ICBs change during the experiment. We found, for both tasks, that the ICBs of each participant in the first and second halves of the task were not significantly different (permutation test identified significant, $p < 0.05$, differences in 14/260 of the pairs, consistent with the null hypothesis), suggesting a limited contribution to ICBs of stochasticity dynamics at these time-scales. It will be interesting to quantify the dynamics of ICBs over longer time-scales.

**The effect of correlations**

In the DDM, spikes are spatially and temporally uncorrelated. As a result, for sufficiently large networks, the magnitudes of both fast and quenched sources of stochasticity decrease as $1/\sqrt{N}$ when $N$ is sufficiently large. Their ratio and hence the distribution of ICBs become independent of $N$ for sufficiently large networks. The two sources of stochasticity also satisfy these scalings in the recurrent spiking model. This is because the network operates in the balanced regime[35,38]. Noise correlations in the spike count of the neurons are therefore very weak[39–41] and the firing rates are widely distributed and are spatially uncorrelated[36].

In case of correlations[40,41], either in the connectivity or in the neuronal activity, spatial heterogeneity and temporal fluctuations decrease inversely proportional to the square root of the *effective* number of independent degrees of freedom. If the effective numbers of degrees of freedom are very different for the quenched and fast noises, one source of stochasticity would dominate, resulting in deterministic or unbiased choices, in contrast to our experimental observations.

**Interpretation of the ICBs observed in our experiments**

Current understanding of the neural mechanisms and computational principles underlying decision-making is based on experiments, in which one manipulates choice preferences of the participants. For example, in operant learning tasks, one investigates how past experience biases choices in favor or against actions that were closely followed by rewards or costs[42–44]. In perceptual tasks, one often studies how specific histories of stimuli bias perceptual decisions[17,45]. Along these lines, it is natural to attribute ICBs to the specific histories of the participants during the experiment. To suppress these effects in the bisection task, vertical impossible trials were always preceded by three horizontal bisection trials. In the motor task, trials were randomized to reduce these effects. To minimize operant learning effects in the bisection task, only delayed and partial feedback was provided. In the motor task we found that dragging time (the feedback) was not substantially different between the first and second motor actions (mean dragging times are $797 \pm 18$ms and $814 \pm 17$ms, respectively; $p > 0.05$, permutation test per participant, not corrected for multiple comparisons) suggesting that feedback played no major role in participants' choices. Combining our experimental results with our modeling work led us to hypothesize that the ICBs we observed are the consequence of random inevitable differences in the fine structure of connectivity in brain areas involved in the decision making process.

**Alternative interpretations and experimental predictions**

In spite of our experimental designs, we cannot exclude the possibility that ICBs that we have observed are the result of sequential or operant effects that we failed to suppress. In particular, ICBs could be the result of such processes which occurred *before* the experiment. For example, considering the impossible trials in our bisection task, participants may prefer to press the Down key because they are accustomed to pressing taskbar icons that are located at the bottom of their computer monitor. Other participants may prefer the Up key because they are used to a taskbar located at the top of the screen. In such a view, ICBs in the vertical bisection task can be attributed to idiosyncratic histories of computer usage prior to the experiment.

Such sequential effects, when analyzed using the DDM, have been interpreted as a manifestation of unbalanced initial conditions, with little effect on the diffusion processes ([46–48], but see[49]). This interpretation is qualitatively different from the one we propose in the present work. In our theoretical models, microscopic heterogeneities in the firing rates (in our DDM) or the connectivity (in our recurrent network model) result in asymmetries in the drift rates.

These two possible interpretations lead to qualitatively different experimental predictions regarding the dependence of the bias on decision threshold. If the bias is due to asymmetric initial conditions, the DDM predicts that increasing the threshold would reduce the bias. This is because the relative difference in the distance of the starting point to the two thresholds, corresponding to the two alternatives, would decrease. By contrast, if the bias is due to asymmetries in the drift rates, increasing the threshold would increase the bias, because a longer integration time allows for the averaging out of the fast noise. Thus, it is in principle possible to experimentally test our hypothesis by investigating how decision time is correlated with the magnitude of ICBs.

There are several pitfalls, however, that should be avoided when investigating such a correlation. First, relating decision times and the distribution of biases in different tasks can be misleading. This is because different tasks may involve different brain areas, in which the levels of fast noise and spatial heterogeneities may differ. Second, comparing the biases of participants with different average decision time performing the same task can also be misleading. This is because differences in average decision time can be due to differences in decision thresholds, but can also be due to different manifestation of the quenched noise. In the former case, average decision time is predicted to be positively correlated with the magnitude of bias. In the latter it would be anti-correlated, the result of the fact that a larger drift in the model is associated with a faster approach to the decision threshold. We intend to develop in future work an experimental paradigm, applicable to a large number of participants, which will allow us to precisely manipulate decision-time in the same task and in the same participant.

**Symmetry and asymmetry in decision tasks**

Both the discrimination and motor tasks that we considered were conceived in an attempt to study symmetric choices. However obviously, completely symmetric tasks exist only in *gedanken* experiments, in which the asymmetry of the task is vanishingly small[1]. In real experiments, *any* two-alternative choice task is inherently asymmetric. At the input level, the two alternatives must sufficiently differ in order to indicate the two possible decisions. At the output level, the two motor actions must differ in order to indicate which decision was made by the participant. These inevitable asymmetries make it impossible to rule out the possibility that ICBs are the result of the interaction of idiosyncratic *history* of the participant with such irreducible asymmetry in the alternative choices or the motor actions used to indicate them.

We argue, however, that even in the ideal limit, in which the task becomes completely symmetric, substantial ICBs remains. In fact, because heterogeneity in firing rates and irregular spike timing are hallmarks of the dynamics of cortical networks[26,50], we argue that the occurrence of an ICB in a cortical-based decision task is almost inevitable. It would therefore be surprising to find a decision task that is devoid of the ICB, unless biases are penalized in the experiment.

**Materials and Methods**

**The perceptual discrimination task**

The study was approved by the Hebrew University Committee for the Use of Human Subjects in Research. Data were collected from 100 participants (51 males, 49 females; 91 dextrals, 7 sinistral, 2 ambidextrous; mean age = 39 years, max = 71 years, min = 22 years). Recruitment was based on the online labor market Amazon Mechanical Turk[51], all of whom are Mechanical Turk's Masters, located in the United States of America. All participants reported normal or corrected to normal vision and no history of neurological disorders. The experiment was described as an academic survey of visual acuity. A base monetary compensation was given to all applied participants for the participation, and additional bonus fee was given according to performance.

*Procedure*

Participants were instructed to indicate the offset direction of the transecting line, out of two alternative responses. Possible responses were either 'left' or 'right', for the horizontal discrimination task, or 'up' or 'down', for the vertical discrimination task. Participants were asked to answer as quickly and accurately as possible.

In each trial, a 200 pixels-long white line, transected by a perpendicular 20 pixels-long white line was presented on a black screen (Fig. 1a). The stimuli were limited to a 400 pixels X 400 pixels square at the center of the screen and window resolution was verified for each participant individually so that it did not exceed the centric box in which all stimuli were presented. The horizontal location of all vertical bisection lines and the vertical location of all horizontal bisection lines were centered. After 1 sec, the stimulus was replaced by a decision screen, composed of two arrows, appearing in opposite sides of the screen, and a middle, 4-squares submit button. The participants indicated their decision by moving the initially centered cursor to one of the arrows, pressing it, and finalizing their decision by pressing the 'submit' button. No feedback was given regarding the correct response; however, the participants were informed about the accumulated bonus fee up to this trial every 30 trials.

The experiment consisted of 240 transected-line trials, 120 horizontal and 120 vertical. Trials were ordered in 80 alternating blocks of 3 horizontal and 3 vertical transected lines. Unbeknown to the participants, there were 20 impossible horizontal and 20 impossible vertical trials (⅙ of the trials). To control for sequential effects, all impossible trials were preceded by trials of the opposite task. The order of the trials was pseudorandom but identical for all participants. For the non-impossible trials, the deviation from the veridical midpoint was uniformly distributed between 5 and 10 pixels, with an equal number of offsets in each direction. To quantify the ICB, we focused on the vertical bisection trials because it is well established that in the horizontal bisection task, participants exhibit a global bias, which has been attributed to pseudoneglect[52].

**The motor task**

The study was approved by the Hebrew University Committee for the Use of Human Subjects in Research. Data were collected from 20 participants (13 males, 7 females; all dextrals; mean age = 25 years, max = 41 years, min = 19 years), who were recruited using on-campus advertising. All participants reported normal or corrected to normal vision and no history of neurological disorders. A base monetary compensation was given to all participants, and additional bonus fee was given according to performance (speed of moving the dots).

*Procedure*

In each trial, a pair of dots, equally distant from a central black disk, were presented on a background of a larger white disk, as depicted in Figs. 2a and S1a. Participants were instructed to drag the two dots into the black disk using the mouse cursor as quickly as possible. Each trial started with a forced delay period of 0.75 seconds. Then, the mouse cursor appeared in the center of the disc. The participant used the mouse to move the cursor to one of the dots. Then, she dragged the chosen dot to the central black disk by pressing the mouse and moving it. If accurate, a release of the dot on the central black disk resulted in a 1.1 sec "swallowing" of the dot animation, indicating a successful drag. The dragging time appeared on the screen (from the time of clicking on the dot to the time of its release) and after the forced delay, it disappeared and the cursor reappeared in the center of the disk. The participant processed the second dot as for the first dot. We used 10 different pairs of dots, each presented 20 times. Each pair of dots was of equal distance from the center of the black disk, but of a different color and a different angular location (Fig. S1b). The order of presentation was pseudorandom such that in every consecutive group of 10 trials all pairs appeared. Participants received a show-up monetary compensation of 20 NIS and received a bonus reward depending on the average dragging duration: 1 NIS for every 50 msec below 1 sec.

**The Drift Diffusion model**

We consider two populations of neurons, denoted by '$U$' and '$D$', representing choice 'Up' and 'Down' (Fig. 3a). Each population is composed of $N/2$ independent Poisson neurons. The stimulus-dependent feedforward inputs to neuron $i$ ($i \in \{1, \ldots, \frac{N}{2}\}$) in population $\alpha$ ($\alpha \in \{U, D\}$) is given by: $\mu_i^\alpha \left(\frac{\Delta L}{L}\right) = k^\alpha \cdot \frac{\Delta L}{L} + z_i^\alpha$, where $k^U = -k^D = k$ is a parameter. The heterogeneity between the neurons due to recurrent input is modelled $z_i^\alpha$, which is independently drawn from a zero-mean Gaussian distribution with variance $\sigma^2$, $\langle z_i^\alpha \rangle = 0$, $\langle (z_i^\alpha)^2 \rangle = \sigma^2$. The firing rate $v_i^\alpha$ of neuron $i$ in population $\alpha$ is given by

$$v_i^\alpha = \bar{v} \cdot e^{\gamma \cdot \mu_i^\alpha \left(\frac{\Delta L}{L}\right)} \quad (1)$$

where $\bar{v}$ is a baseline firing rate and $\gamma$ is the gain[36]. Note that due to the exponential transfer function and the normal distribution of inputs, the firing rates are log-normally distributed, as reported in the cortex[25,26]. The dynamics of choice follows DDM, as described in the text[27]. In particular, a decision is made when $|\Delta n|$, reaches for the first time a given threshold, $\theta = \sqrt{N} \cdot \tilde{\theta}$, where $\tilde{\theta}$ is a parameter of the model, which is $O(1)$.

For $N \gg 1$, neglecting the threshold effect, the difference in spike count is given by $\Delta n(t) \sim \mathcal{N}(\Delta v \cdot t, \Sigma v \cdot t)$, where $\Delta v = \sum_i v_i^U - \sum_i v_i^D$ and $\Sigma v = \sum_i v_i^U + \sum_i v_i^D$. For $N \gg 1$, both $\Delta v$ and $\Sigma v$ are normally distributed:

$$\Delta v \sim \mathcal{N}\left(N\bar{v} e^{\frac{\gamma^2 \sigma^2}{2}} \sinh\left(\gamma k \frac{\Delta L}{L}\right), N\bar{v}^2 (e^{\gamma^2 \sigma^2} - 1) e^{\gamma^2 \sigma^2} \cosh\left(2\gamma k \frac{\Delta L}{L}\right)\right)$$

$$\Sigma v \sim \mathcal{N}\left(N\bar{v} e^{\frac{\gamma^2 \sigma^2}{2}} \cosh\left(\gamma k \frac{\Delta L}{L}\right), N\bar{v}^2 (e^{\gamma^2 \sigma^2} - 1) e^{\gamma^2 \sigma^2} \cosh\left(2\gamma k \frac{\Delta L}{L}\right)\right).$$

Note that $\Delta n$ and $\Delta v$ correspond to different stochastic processes: the stochasticity of $\Delta n$ stems from trial-by-trial variability, conditioned on the firing rates of the neurons. By contrast, the stochasticity of $\Delta v$ reflects heterogeneity in these firings rates across different realizations of decision-making networks.

The standard deviation of the distribution of $\Sigma v$ is of $O(\sqrt{N})$, whereas its mean is $O(N)$ even when $\Delta L \to 0$. Therefore, in the limit $N \gg 1$, $\Sigma v \approx N\bar{v}e^{\frac{\gamma^2\sigma^2}{2}}\cosh\left(\gamma k \frac{\Delta L}{L}\right)$. By contrast, in the regime in which $\left|\frac{\Delta L}{L}\right| = O(1/\sqrt{N})$, the mean and standard deviations of the distribution of $\Delta v$ are comparable, both are $O(\sqrt{N})$.

The probability of an 'Up' decision is obtained by solving a first-passage problem:

$$p \equiv \Pr('\text{Up}') = \left(1 + e^{-2\Delta v \sqrt{N} \cdot \tilde{\theta}/\Sigma v}\right)^{-1} = \left(1 + e^{-2\frac{\Delta v}{\sqrt{N}} \cdot \frac{\tilde{\theta}}{\bar{v} \cdot \cosh\left(\gamma k \frac{\Delta L}{L}\right) \cdot e^{\frac{\gamma^2 \sigma^2}{2}}}}\right)^{-1} \quad (2)$$

The psychometric curve is obtained by substituting the dependence of $\Delta v$ on $\Delta L$. Specifically, when the two networks are symmetric, $\Delta v = N\bar{v}e^{\frac{\gamma^2\sigma^2}{2}}\sinh\left(\gamma k \frac{\Delta L}{L}\right)$, yielding:

$$p = \left(1 + e^{-2\sqrt{N}\cdot\tilde{\theta}\tanh\left(\gamma k \frac{\Delta L}{L}\right)}\right)^{-1} \approx \left(1 + e^{-2\sqrt{N}\cdot\tilde{\theta}\cdot\gamma k \cdot \frac{\Delta L}{L}}\right)^{-1}$$

More generally, when the two networks are only drawn from the same distribution, the resultant psychometric curve will be horizontally shifted relative to the identical networks case.

To compute the distribution of choice biases, we consider the case in which the external input is symmetric, $\Delta L = 0$ and thus $\Delta \nu \sim \mathcal{N}\left(0, N\bar{\nu}^2\left(e^{\gamma^2\sigma^2} - 1\right) \cdot e^{\gamma^2\sigma^2}\right)$. Using the change-of-variable technique,

$$\Pr(p) = \frac{1}{p \cdot (1-p)} \cdot \frac{1}{\sqrt{8\pi\left(e^{\gamma^2\sigma^2}-1\right)} \cdot \tilde{\theta}} e^{-\frac{(\log(p)-\log(1-p))^2}{8\left(e^{\gamma^2\sigma^2}-1\right)\tilde{\theta}^2}} \quad (3)$$

Choice bias in this framework is given by $2p - 1$.

The corresponding distribution of decision times is computed by averaging the drift-conditioned distribution of first-passage times over the distribution of $\Delta\nu$, yielding[28,53,54]:

$$f(t) = \frac{\pi}{2\theta^2} \bar{\nu} e^{\frac{\gamma^2\sigma^2}{2}} \frac{1}{\sqrt{1+t\bar{\nu}e^{\frac{\gamma^2\sigma^2}{2}}(e^{\gamma^2\sigma^2}-1)}} \exp\left(\frac{1}{2} \frac{\tilde{\theta}^2}{t\bar{\nu}e^{\frac{\gamma^2\sigma^2}{2}} + \frac{1}{(e^{\gamma^2\sigma^2}-1)}}\right)$$

$$\times \sum_{k=1}^{\infty} k \sin\left(\frac{\pi k}{2}\right) \exp\left(-t \frac{k^2\pi^2}{8} \frac{\bar{\nu}}{\tilde{\theta}^2} e^{\frac{\gamma^2\sigma^2}{2}}\right) \quad (4)$$

**The spiking network model**

The model consists of a recurrent network of $N$ leaky-integrate-and-fire (LIF) neurons, $N^E = 0.8N$ excitatory and $N^I = 0.2N$ inhibitory (the superscript notation here denotes neuron type, excitatory or inhibitory, rather than the selectivity of the neuron).

*Single neuron dynamics:* The sub-threshold dynamics of the membrane potential, $V_i^\alpha(t)$, of neuron $i$ in population $\alpha$ ($i = 1, \ldots, N^\alpha$; $\alpha = E, I$) follow:

$$\tau_m \frac{dV_i^\alpha(t)}{dt} = -(V_i^\alpha(t) - V_L) + I_{rec,i}^\alpha(t) + I_{FF,i}^\alpha(t) + I_b^\alpha$$

where $\tau_m$ is the neuron membrane time constant, $V_L$ is the reversal potential of the leak current. Inputs to the neuron are modeled as currents: $I^\alpha_{rec,i}(t)$ is the recurrent input into neuron $(i,\alpha)$, due to its interactions with other neurons in the network, $I^\alpha_{FF,i}(t)$ is the feedforward input into that neuron elicited upon presentation of the stimulus, and $I^\alpha_b$ is a background feedforward input, independent of the stimulus, identical for all the neurons and constant in time. These subthreshold dynamics are supplemented by a reset condition: if at $t = t^\alpha_i$ the membrane potential of neuron $(i,\alpha)$ reaches the threshold, $V^\alpha_i(t^{\alpha-}_i) = V_T$, the neuron fires an action potential and its voltage resets to $V^\alpha_i(t^{\alpha+}_i) = V_R$.

*The feedforward input:* The network consists of two types of neurons, *U*-selective and *D*-selective. In the absence of stimulus, the feedforward input $I^\alpha_{FF,i}(t) = 0$ for all the neurons. Upon presentation of a stimulus $I^\alpha_{FF,i}(t)$ to *U*-selective neurons is stronger when the upper segment is larger than the lower one. It is the opposite for the *D*-selective neurons. Specifically, we take:

$$I^\alpha_{FF,i}(t) = I^\alpha_0 + \varepsilon \frac{\Delta L}{L} I^\alpha_1$$

where $L$ is the length of the line, $\Delta L$ is the difference between the length of the upper segment and the lower segment, $I^\alpha_0$ and $I^\alpha_1$ are positive constants and $\varepsilon$ denotes the selectivity type of the neuron: $\varepsilon = +1$ for *U* neurons and $\varepsilon = -1$ for *D* neurons. Thus, the *U* (*D*) neurons are selective to stimulus in which the upper segment is longer (shorter) than the lower segment. We denote the set of *U*-selective (*D*-selective) neurons in population $\alpha = E, I$ by $U^\alpha$ ($D^\alpha$). Neuron $(i,\alpha) \in U^\alpha$ when $i = 1 \ldots \frac{N^\alpha}{2}$ and t $(i,\alpha) \in D^\alpha$ when $i = \frac{N^\alpha}{2} + 1 \ldots N^\alpha$.

*The recurrent input:* The connectivity matrix is composed of two components: the non-specific (independent of the selectivity of the pre and postsynaptic neurons) component is fully random (Erdös-Renyi graph). The corresponding $N^\alpha \times N^\beta$ connectivity matrix, $\mathbf{C}_{NS}^{\alpha\beta}$, is such that $C_{NS,ij}^{\alpha\beta} = 1$ with probability $K/N^\beta$ and $C_{NS,ij}^{\alpha\beta} = 0$ otherwise, where $K$ is the average number of non-specific inputs that a neuron receives from neurons in population $\beta$. The strength of the non-specific connections depends solely on $\alpha, \beta$ yielding: $J_{NS,ij}^{\alpha\beta} = J_{NS}^{\alpha\beta} C_{NS,ij}^{\alpha\beta}$ where $J_{NS}^{\alpha E}>0$ (excitation) and $J_{NS}^{\alpha I}<0$ (inhibition).

The competition between the *U* and the *D* selective neurons is mediated by an additional set of connections, which are specific. These connections are much less numerous but stronger than the unspecific ones. The corresponding connectivity matrices, $\mathbf{C}_{S,ij}^{\alpha\beta}$, are such that:

1) $C_{S,ij}^{\alpha E} = 0$ i.e. we assume no specific excitation.
2) $C_{S,ij}^{\alpha I} = 0$ if $i$ and $j$ have the same selectivity properties.
3) $C_{S,ij}^{\alpha I} = 1$ with probability $2\sqrt{K}/N^I$ if $i$ and $j$ have different selectivity properties.

Therefore, each neuron receives, on average, $\sqrt{K}$ connections from inhibitory neurons whose selectivity is different from its own (compared with, on average, $K$ non-selective inhibitory connections).

The strength of the specific connections depends solely on the neurons' type $J_{S,ij}^{\alpha I} = J_S^{\alpha I} C_{S,ij}^{\alpha I}$; $g = J_S^{\alpha I}/J_{NS}^{\alpha I}$.

The total current into neuron $(i, \alpha)$ due to the recurrent interactions is:

$$I^{\alpha}_{rec,i}(t) = \sum_{j,\beta} \left(J^{\alpha\beta}_{S,ij} + J^{\alpha\beta}_{NS,ij}\right) S^{\alpha\beta}_j(t)$$

where $S^{\alpha\beta}_i(t)$ are synaptic variables, which follow the dynamics:

$$\tau_S \frac{dS^{\alpha\beta}_i(t)}{dt} = -S^{\alpha\beta}_i(t) + \sum_{\{t^{\beta}_j\}} \delta(t - t^{\beta}_j)$$

Here $\tau_S$ is the synaptic time constant (assumed to be the same for all synapses) and the sum is over all spikes emitted at times $t^{\beta}_j < t$.

*Decision-making and decision criterion:* In response to the presentation of a stimulus, the activities of the *U*-selective and *D*-selective neurons change differently (Fig. 5d). We compute at all time points the population averaged activity of all the excitatory neurons which belong to the set $a$, ($a$-selevtive), denoted by $\nu_a$, $a \in \{U, D\}$, by convolving the spike times with an exponential filter with a time constant of 50 msec. Decision is based on the ratio: $\frac{\nu_U - \nu_D}{\nu_U + \nu_D}$. If $\frac{\nu_U - \nu_D}{\nu_U + \nu_D} > \phi$, the decision provided by the network is that upper segment is longer than the lower one, whereas for $\frac{\nu_D - \nu_U}{\nu_D + \nu_U} > \phi$ it is the opposite, where $\phi > 0$ is a parameter.

It should be noted that the ability of the network to make a decision depends on the network parameters, in particular on $g$, which controls the relative strength of the competition between *U* and *D* neurons, the value chosen for the threshold $\phi$ as well as the parameters of the stimulus, $\bar{I}^{\alpha}_0$ and $\bar{I}^{\alpha}_1$. An extensive study of this issue is beyond the scope of the present paper. We have chosen all the parameters such that the network is always able to make decisions, including when $\Delta L = 0$

*Numerical integration:* The dynamics of the model circuit were numerically integrated using the Euler method supplemented with an interpolation estimate of the spike times[55]. In all simulations the integration time step was 0.1 msec. We verified the validity of the results by performing complementary simulations with smaller time steps.

*Model parameters*: The parameters used in all the simulations were: $V_L = -60\text{mV}$; $\tau_m = 10\text{msec}$; $V_T = 10\text{mV}$; $V_R = -60\text{mV}$; $J_{EE} = 35\text{mV}\cdot\text{ms}, J_{IE} = 233.3\text{mV}\cdot\text{ms}, J_{EI} = -175\text{mV}\cdot\text{ms}, J_{II} = -233.3\text{mV}\cdot\text{ms}, I_b^E = 840\text{mV}, I_b^I = 560\text{mV}, I_0^E = 840\text{mV}, I_0^I = 560\text{mV}, I_1^E = 140\text{mV}, I_1^I = 140\text{mV}, \tau_S = 3\text{msec}$. The total number of neurons and average non-specific connectivity were $N = 40{,}000$, $K = 400$, $\phi = 0.4$.

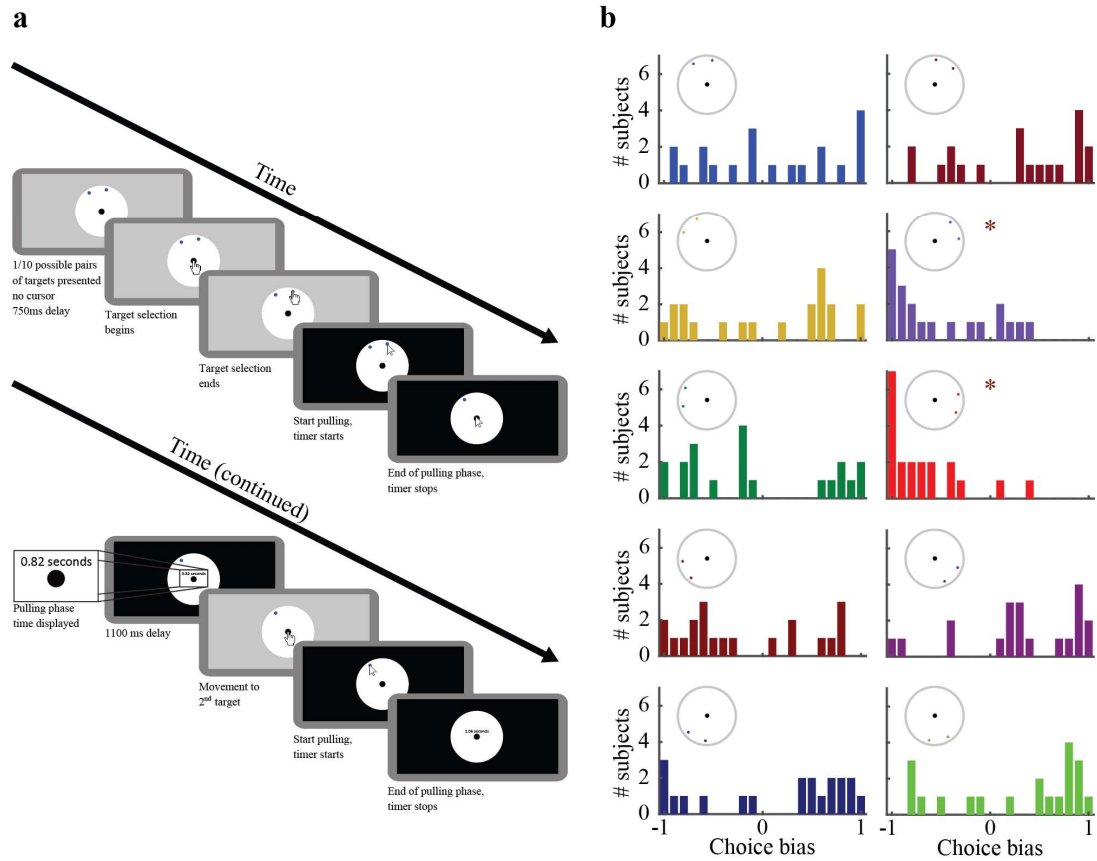

**Figure S1** The motor task. (a) A detailed description of a single trial in the motor task. A trial began with a presentation of the two dots and a central black disk on a white disk background. After 750 msec, a cursor appeared on the central black disk and the participant used the mouse to place it over one of the dots. Then, the participant clicked on the dot and used the mouse to pull it to the central black disk and released it. An accurate release of the dot resulted in a 1.1 sec "swallowing" of the dot animation the dragging time (from click to release) was displayed (participants were instructed to minimize that time). Then, the cursor reappeared on the central black disk and the process repeated. (b) Locations and colors of all 10 pairs of dots used in the experiment, with the corresponding distributions of choice biases. There was a consistent choice preference for the two pairs of dots denoted by asterisks and therefore they were not used in the analysis in Fig. 2b.

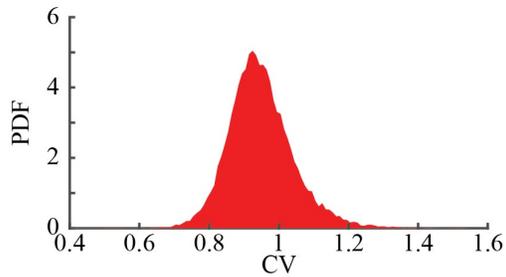

**Figure S2** The distribution of coefficients of variation (CV) of the excitatory neurons in the spiking network model. The CV's were computed over 100 sec of spontaneous activity.

## Data availability

The data that support these findings are available from the corresponding author upon request.

## Acknowledgements

We thank Gianluigi Mongillo, Haim Sompolinsky and Talia Tron for discussions. This work was conducted within the scope of the France-Israel Laboratory of Neuroscience. D. H. thanks the Department of Neurobiology at the Hebrew University for its warm hospitality. This work was supported by the Israel Science Foundation (Y. LO., Grant No. 757/16), the DFG (Y. LO.), the Gatsby Charitable Foundation (Y. LO.), ANR-09-SYSC-002-01 (D. H.) and the France-Israel High Council for Science and Technology (D. H. and Y. LO.).

## Author contributions

L. L., Y. LA., D. H. and Y. LO. conceived and planned the experiments; L. L., R. D., D. H. and Y. LO. developed the models; L. L., D. H. and Y. LO. wrote the manuscript.